\documentclass[conference,compsoc]{IEEEtran}
\IEEEoverridecommandlockouts 
\ifCLASSOPTIONcompsoc
  \usepackage[nocompress]{cite}
\else
  \usepackage{cite}
\fi
\ifCLASSINFOpdf
  \usepackage[pdftex]{graphicx}
  \DeclareGraphicsExtensions{.pdf,.jpeg,.png}
\else
\fi
\usepackage{tikz}
\usetikzlibrary{shapes.geometric, arrows}
\usepackage{amsmath}

\usepackage{algorithm}
\usepackage{algorithmic}
\usepackage{amsfonts} 

\usepackage{booktabs}

\usepackage{url}
\usepackage{xurl} 
\usepackage{hyperref} 

\usepackage[utf8]{inputenc}

\begin{document}
%
\title{zk-ScalHard: Scalable and Hardware-Rooted Proactive Attestation \\for Secure Pull-based OTA Updates in Zonal SDVs}



\author{
    \IEEEauthorblockN{
        Shrikant Tangade\textsuperscript{1,2*}, 
        Bansi Pambhar\textsuperscript{1}, 
        Valeria Loscrì\textsuperscript{3}, 
        and Mauro Conti\textsuperscript{4,5}
    }
    \IEEEauthorblockA{\textsuperscript{1}autoMoTIVe-X Lab, Belagavi, India}
    \IEEEauthorblockA{\textsuperscript{2}Anuvartik Mirji Bharatesh Institute of Technology, Belagavi, India}
    \IEEEauthorblockA{\textsuperscript{3}SERENDIPITY Team, Inria Lille -- Nord Europe, France}
    \IEEEauthorblockA{\textsuperscript{4}University of Padua, Padua, Italy}
    \IEEEauthorblockA{\textsuperscript{5}Örebro University, Örebro, Sweden}
    \IEEEauthorblockA{Emails: \{shrikant.tangade, bansi.pambhar\}@motivex.in, shrikant.tangade@ambit.edu.in, valeria.loscri@inria.fr, \\ mauro.conti@unipd.it}
    
    \thanks{*Corresponding author: shrikant.tangade@motivex.in. S. Tangade is an MSCA SoE Fellow at Inria Lille, Director of the autoMoTIVe-X Lab, and a Professor at AMBIT, Belagavi, India. B. Pambhar is a Research Assistant at the autoMoTIVe-X Lab. V. Loscrì is the Head of the SERENDIPITY Team at Inria Lille. M. Conti is a Professor with the University of Padua and Örebro University, and Head of the SPRITZ Research Group. This work was supported by the MSCA Seal of Excellence (SoE), awarded by the European Commission and funded by the University of Lille. Source code: \url{https://anonymous.4open.science/r/zk-ScalHard-8C52/README.md}.
    }
}



%



\maketitle


\begin{abstract}
Today, software-defined vehicles (SDVs) are transitioning to zonal-oriented architecture (ZoA), while the zero-day vulnerability window is collapsing from an average of 771 days to less than 60 minutes. This shift necessitates a paradigm move from reactive, cloud-initiated, 'push-based' updates to proactive, vehicle-initiated, 'pull-based' Over-the-Air (OTA) update lifecycles. Although pull-based OTA updates enhance proactive security, requiring 100+ electronic control units (ECUs) to perform frequent firmware state attestation creates a critical scalability crisis. Current industry-standard OTA updates, such as Uptane and AUTOSAR Adaptive, rely on PKI-based models that exhibit unsustainable linear $O(n)$ overhead, leading to significant bandwidth bottlenecks and the exposure of sensitive configuration metadata. To address these challenges, we propose {\itshape zk-ScalHard}, a scalable and privacy-preserving silicon-to-cloud authentication protocol. {\itshape zk-ScalHard} provides a hierarchical trust-promotion model for the attestation handshake in pull-based OTA updates without replacing existing firmware delivery mechanisms. We integrate hardware-rooted Silicon Physical Unclonable Functions (PUFs) with BCH-based fuzzy extractors to ensure stable, deterministic witnesses without persistent key storage. We design and implement two zero-knowledge proof (ZKP) circuits: (1) zonal identity and integrity (ZIDI) and (2) high-performance computing aggregation (HPCA). By utilizing in-situ multi-party computation (MPC) in ZIDI, we decentralize trust initialization to the vehicle level, while recursive proof aggregation in HPCA enables constant-size vehicle-to-cloud (V2I) proofs. Benchmarks against the industry-standard Uptane framework demonstrate that {\itshape zk-ScalHard} achieves constant $O(1)$ communication and verification complexity, down from linear $O(n)$. Evaluation results show a 99.2\% reduction in V2I bandwidth and a 99.9\% reduction in the temporal attack surface for key extraction. These results demonstrate that {\itshape zk-ScalHard} provides deterministic scalability and 100\% vehicle data sovereignty, aligning with GDPR compliance required for the proactive security of next-generation Zonal SDV fleets.
\end{abstract}


%
\IEEEpeerreviewmaketitle

\section{Introduction}

The next-generation automotive industry is moving towards software-defined vehicles (SDVs). Advanced SDVs equipped with 100+ electronic control units (ECUs) are also called computers-on-wheels. Each ECU performs its task, such as automatic braking or steering control, using specific software/ firmware. This software enables improvements in vehicle efficiency, new feature additions, and bug fixes through over-the-air (OTA) updates with the Internet, rather than through traditional offline updates (i.e., taking the vehicle to the service station). 
In 2024 (Q4), 5.1 million Tesla vehicles received bug fixes via OTA updates \cite{hinton}. The OTA updates benefit both automotive original equipment manufacturers (OEMs) and vehicle owners. The OEMs save intermediate vehicle service stations billions of dollars on software updates. In addition, OEMs are adopting a software-as-a-service (SaaS) business model to drive revenue. Whereas vehicle owners save time by not taking their vehicles to a service station and receive updates from their home or office parking. Furthermore, vehicles are now more customizable and have longer lifespans thanks to software-driven OTA updates.

The OTA update process connects SDV's 100+ ECUs to the OEM's backend OTA cloud server through the Internet. This introduces a new cyberattack surface that hackers can exploit, enabling them to inject malicious software into ECUs and take control of the vehicle. This leads to safety-critical issues in which passengers' lives and the vehicle itself are at risk. To ensure SDVs' safety and security, the United Nations (UN) introduced new automotive-specific regulations, UN Regulation No. 155 (UN-R155) \cite{unr155} and UN Regulation No. 156 (UN-R156) \cite{unr156}, for cybersecurity and software updates, respectively. The automotive industry introduced two international standards for road vehicles to comply with these regulations: ISO/SAE 21434 - the cybersecurity management system (CSMS) \cite{iso21434}, which aligns with UNR155 and ISO/SAE 24089 - the software update management system (SUMS) \cite{iso24089}, which aligns with UNR156. To implement SDV’s robust security and enhanced safety, these regulations and standards must complement each other. This makes {\itshape secure} OTA updates mandatory for SDV, not optional \cite{unecer155156}.

Since SDVs are connected to the external world via wireless networks such as Wi-Fi and the Internet, they are more prone to remote cyberattacks. This forces frequent OTA updates to fix bugs and also enable/add new features based on customer needs. When the OTA update system itself is insecure, it exposes the system to serious security threats, such as malware injection and remote vehicle control. In today’s transitional automotive industry, the supply chain (i.e., OEMs, Tier 1, \& Tier 2 suppliers) is facing challenges in implementing a {\itshape secure} OTA update system to ensure SDVs' safety. In addition, it should comply with new automotive {\itshape cybersecurity regulations/standards} (UNR155, R155; ISO/SAE 21434, 24089) and general-purpose {\itshape data regulations} such as the GDPR, CCPA, and the EU Cybersecurity Act to ensure SDVs' data privacy and sovereignty. We define data sovereignty in the context of secure OTA updates for Zonal SDVs as the vehicle's architectural capability to maintain exclusive control over its sensitive configuration metadata and physical identity. 

 

\subsection{Motivation}
Next-generation SDVs are increasingly characterized as connected and autonomous vehicles (CAVs). This transition marks the rise of the 'Data Centre on Wheels' \cite{mckinsey2019}, where the vehicle's value is increasingly defined by its software stack and cloud connectivity. Further, the automotive Electric \& Electronics (E/E) architecture is evolving from legacy centralized gateway architecture (CGA) and domain-oriented architecture (DoA) toward zonal-oriented architecture (ZoA) \cite{pierre24}. Recent automotive industry projections estimate that adoption will rise from 2\% in 2023 to nearly 45\% in ZoA in 2030 \cite{Sebastian25} with 80\% of OEMs already initiating this structural shift \cite{Harsha26}. In parallel with this architectural evolution, communication technologies are migrating from Controller Area Network (CAN) to Automotive Ethernet (AE), facilitating the shift from traditional offline maintenance to frequent OTA updates. 

While CGA and DoA models rely on relatively static configurations with limited OTA capabilities, the ZoA supports dynamic orchestration and frequent software deployment for 100+ ECUs. The current focus in the automotive industry remains on “pull-based” OTA updates, in which the vehicle initiates the update request. These pull-based OTA updates require a robust authentication mechanism to verify the vehicle’s identity and software metadata against the OTA server before releasing sensitive firmware binaries.

 However, the shift to Zonal SDVs introduces significant security challenges. Supporting flexible OTA updates for high-density ECU environments creates a massive scalability bottleneck and exposes sensitive configuration metadata, compromising data privacy. Furthermore, the hardware heterogeneity between resource-constrained edge ECUs and the central HPC introduces systemic vulnerabilities and potential single points of failure. Consequently, there is an urgent need for a Zonal-native, scalable, and decentralized authentication protocol that preserves privacy throughout the secure OTA update lifecycle.

\subsection{Problem Statement and Gap Analysis}
Current industry standards for OTA updates, such as the Uptane framework and the AUTOSAR Adaptive Platform, provide robust integrity but face a critical “Scalability Wall” in terms of bandwidth, latency, and data privacy. When an SDV initiates a software update request, it must transmit a comprehensive vehicle manifest containing metadata for 100+ ECUs. This results in unsustainable communication and computational overheads that act as bottlenecks for high-density zonal networks.

To quantify the scalability limits of these standards, we conducted a systematic benchmarking of the Uptane reference implementation (detailed in Section 7 \& Appendix). Our evaluation reveals that, for an SDV equipped with 100 ECUs, the vehicle version manifest (VVM) payload grows linearly ($O(n)$) to 98.6KB. Furthermore, the total VVM verification latency reaches 15.4s per vehicle. Beyond performance, these frameworks rely on centralized, trusted authorities (such as PKI) for key management and store persistent secrets in Flash memory, creating a systemic single point of failure. While existing scientific research has addressed individual aspects of authenticity, scalability, and privacy, no current solution simultaneously meets all these requirements for vehicle authentication during OTA updates (see Section 9: Related Work).

To resolve these conflicts, we introduce {\itshape zk-ScalHard}, a scalable authentication protocol built on a multi-tier, heterogeneous trust-promotion architecture. This model maps cryptographic intensity to the underlying hardware constraints of Zonal SDVs, replacing stateful, identity-revealing signatures with hardware-bound, recursive proofs. By moving the “Trust Ceremony” from the cloud to the vehicle’s Zonal Controllers, we eliminate centralized risk while achieving asymptotic $O(1)$ scalability.

Table \ref{tab:compare} provides a comparative gap analysis between zk-ScalHard and current industry and research standards. While legacy frameworks like Uptane and emerging PQC standards address integrity and quantum threats, they suffer from linear metadata bloat ($O(n)$) and persistent exposure of secrets.

\begin{table*} [t]
\centering
  \caption{Comparative Gap Analysis of Automotive Authentication Solutions}
  \label{tab:compare}
  \begin{tabular}{lccccc}
    \toprule
    \textbf{Framework} &\textbf{Asymptotic Scalability}& \textbf{Temporal Isolation}& \textbf{Identity Privacy}& \textbf{Trust Model}\\
    \midrule
   \textbf{Uptane/ PKI} \cite{8Ref48,9Ref49} & Linear $O(n)$ & No (Persistent)& Revealed& Centralized\\
   \\
   \textbf{IBM Idemix} \cite{15Ref56,16Ref57}& Linear $O(n)$& No (Persistent)& \textbf{Zero-Knowledge}& Centralized \\
   \\
   \textbf{MS U-Prove} \cite{17Ref60}& Linear $O(n)$  & No (Persistent) & Selective & Centralized\\
   \\
   \textbf{NIST PQC} \cite{20Ref42,21Ref40}& Linear $O(n)$& No (Persistent) & Revealed & Centralized\\
   \\
\textbf{zk-ScalHard} (Proposed)& \textbf{Constant $O(1)$} & \textbf{Yes (Ephemeral)}  & \textbf{Zero-Knowledge} & \textbf{Decentralized}\\
\\

    \bottomrule
  \end{tabular}
\end{table*}

zk-ScalHard is the first to achieve asymptotic invariance ($O(1)$) in both communication and verification, while simultaneously providing temporal isolation to protect physical silicon identities.

\subsection{Research Questions}
To secure the next generation Zonal SDVs OTA updates and compliance with new automotive cybersecurity regulations/ standards, this paper addresses the following four fundamental research questions (RQ):
\begin{itemize}
\item RQ1 ({\itshape Trust Architecture}): How can a multi-layer trust hierarchy be designed to bridge the gap between a resource-constrained ECU (Layer-0) and a global OTA cloud server (Layer-3) while aligning cryptographic intensity with heterogeneous hardware capabilities?

\item RQ2 ({\itshape Asymptotic Scalability}): Is it possible to achieve constant communication and verification complexity ($O(1)$) as ECU density increases, thereby decoupling security overhead from the Zonal SDV architectural scale?

\item RQ3 ({\itshape Decentralized Trust}): How can a Zonal SDV's cryptographic root-of-trust be initialized via in-situ multi-party computation (MPC) to eliminate the systemic single point of failure inherent in centralized trusted setups?

\item RQ4 ({\itshape Temporal Isolation}): How can the temporal attack surface for key extraction be minimized by integrating silicon PUFs as dynamic witnesses, thereby shifting from persistent stateful storage to a transient execution window?  
\end{itemize}

\subsection{Our Contributions}
We address the above challenges and research questions by designing and implementing a novel zonal SDV native hybrid, heterogeneous trust-promotion hierarchy {\itshape zk-DieHard-SDV} authentication protocol for secure OTA updates. Our key contributions to advancing the state of the art are as follows:

\begin{itemize}
\item Contribution 1 ({\itshape Hierarchical Trust Architecture}): We propose {\itshape zk-ScalHard}, a multi-tier hierarchical architecture for trust promotion. It orchestrates trust from edge ECUs to the central OTA cloud server by integrating hardware-rooted silicon PUFs as a temporal dynamic witness (Layer-0), performing in-situ decentralized trust ceremonies (Layer-1), and executing recursive proof aggregation (Layer-2). This architecture achieves scalable, secure, and GDPR-compliant vehicle-to-cloud (Layer-3) attestation within the OTA update lifecycle.  

\item Contribution 2 ({\itshape Asymptotic Scalability}): We design and implement a High-Performance Computing Aggregation (HPCA) ZKP circuit that performs recursive proof aggregation at the vehicle central hub (Layer-2). By compressing four zonal proofs into a single atomic vehicle-level proof for the cloud (Layer-3), we achieve constant-size $O(1)$ communication and verification complexity in contrast to the linear $O(n)$ growth of the industry-standard Uptane framework. Our evaluation quantifies that for a Zonal SDV with 100 ECUs, {\itshape zk-ScalHard} reduces V2I bandwidth by 99.2\% (from 98.6KB to 809B) and achieves a constant V2I verification latency of 2.1s (down from 15.47s), representing a 7.3x speedup over the Uptane framework. This effectively decouples security overhead from Zonal SDV architectural scale.

\item Contribution 3 ({\itshape Decentralized Trust}): We design the Zonal Identity and Integrity (ZIDI) circuit to enable an in-situ zonal multi-party computation (MPC) ceremony at each Zonal Central Unit-ZCU (Layer-1). By using noisy silicon entropy from local edge ECUs to generate unique ZKP proving ($pk$) and verification ($vk$) keys locally, we eliminate the systemic single point of failure inherent in centralized OTA-trusted setups. This anchors the cryptographic root of trust within the vehicle’s physical zonal topology, providing resilience against global parameter compromise in the OTA cloud.

\item Contribution 4 ({\itshape Temporal Isolation}): We propose a stateless authentication model that integrates silicon PUFs as dynamic witnesses across the vehicle hierarchy (ECUs, ZCUs, and HPC). These physical secrets are generated on-the-fly and reside in volatile memory for only 4.2s during the ZKP generation window. Our evaluation demonstrates a 99.9\% reduction in the temporal attack surface for physical key extraction. This model, combined with the ZIDI and HPCA circuits, ensures 100\% vehicle-level data sovereignty by eliminating the transmission of sensitive identity metadata to the OTA cloud. 
\end{itemize}

\section{Background}
To contextualize the design and evaluation of \emph{zk-ScalHard}, this section delineates the fundamental technological and cryptographic pillars upon which our hierarchical protocol is built. We first describe the Zonal SDV architecture to establish the physical network constraints and trust boundaries of modern vehicles. We then provide an overview of the industry-standard Uptane framework to highlight the specific scalability and privacy gaps our work aims to address. Finally, we formalize the Zero-Knowledge Proof (ZKP) primitives, specifically Groth16 and Plonky3, which serve as the mathematical engines for our multi-tier trust promotion model.
\subsection{SDV Zonal Architecture}
\begin{figure}[ht]
  \centering
  \includegraphics[width=\linewidth]{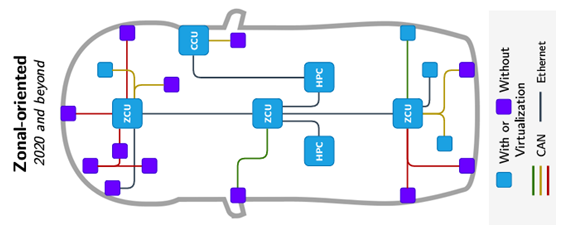}
  \caption{SDV's Zonal-oriented Architecture (ZoA)}
  \label{fig:zonal_sdv}
\end{figure}

The SDV zonal architecture is a centralised system in which edge ECUs are grouped into 3 to 5 physical zones (e.g., front-left, front-right, centre) as shown in Fig. \ref{fig:zonal_sdv}. These are managed by zonal control units (ZCUs) through limited-bandwidth communication protocols such as CAN, CAN-FD, LIN, and FlexRay. These ZCUs connect to central 2-3 high-performance Computing (HPC) systems via the high-speed AE protocol. The SDV's HPC connects to the OEM's OTA update cloud server over the internet to receive frequent software updates.

\subsection{Uptane Framework and OTA Update}
\begin{figure}[t]
  \centering
  \includegraphics[width=\linewidth]{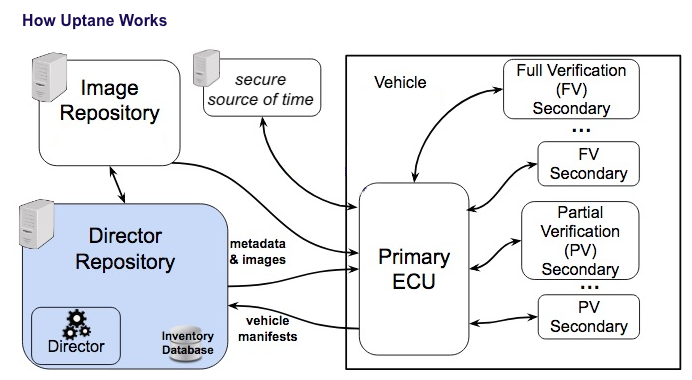}
  \caption{Uptane Secure OTA Update Framework Design (Adapted from \cite{uptane_design})}
  \label{fig:uptane_design}
\end{figure}

Uptane is the current industry standard for secure Over-the-Air (OTA) updates, extending The Update Framework (TUF) to the automotive context. As illustrated in Fig. \ref{fig:uptane_design} \cite{uptane_design}, it utilizes a 'Separation of Trust' model involving four primary roles: the Image Repository, the Director Repository, the Primary ECU (HPC), and Secondary ECUs. Security is maintained through Public Key Infrastructure (PKI), where each ECU generates an ECU Version Report (EVR) signed with an asymmetric key. The Primary ECU aggregates these into a Vehicle Version Manifest (VVM). However, this model introduces a linear $O(n)$ metadata dependency; as ECU density increases, the VVM size balloons, leading to significant bandwidth bottlenecks and increased verification latency at the cloud tier.

\subsubsection{Operational Workflow and Metadata Lifecycle}

\begin{figure}[t]
  \centering
  \includegraphics[width=\linewidth]{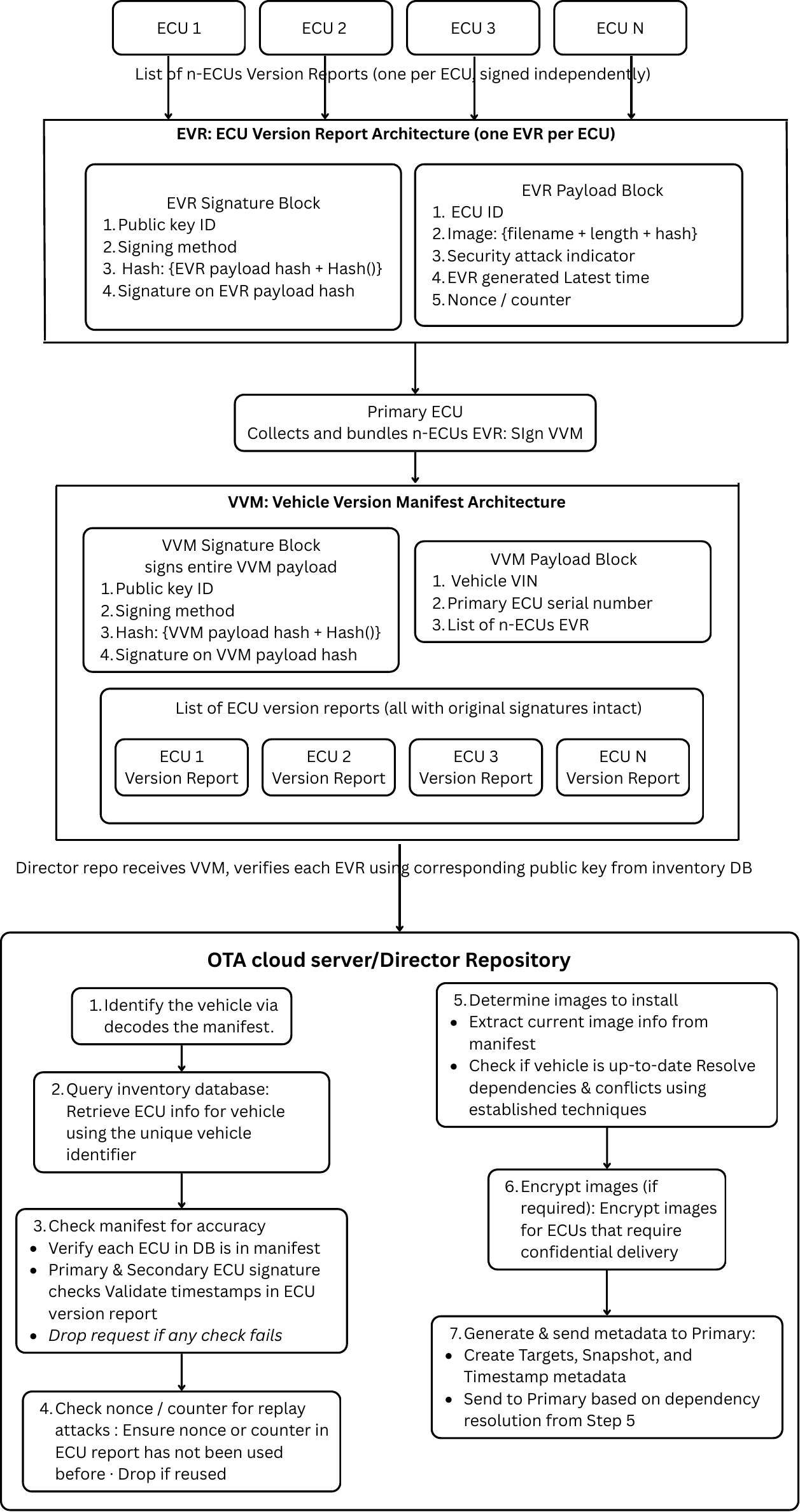}
  \caption{Uptane Operational Workflow and Metadata Lifecycle. The diagram illustrates the linear accumulation of $n$ ECU Version Reports (EVRs) into a single Vehicle Version Manifest (VVM), highlighting the $O(n)$ signature-verification bottleneck at the Cloud tier.}
  \label{fig:uptane_workflow}
\end{figure}

The end-to-end update lifecycle and metadata orchestration within the Uptane framework are illustrated in Fig. \ref{fig:uptane_workflow}. Algorithm \ref{alg:uptane_flow} formalizes the high-level operational flow, highlighting the sequential accumulation of individual ECU signatures. The process begins with an OTA update ‘pull’ request at each edge EUC, followed by the generation of an ECU Version Report (EVR). Each ECU’s EVR contains its current firmware hash with a digital signature. As illustrated, the Primary ECU (HPC) collects and bundles $n$ individual EVRs into a comprehensive VVM.

This architecture reveals the fundamental scalability flaw of legacy standards: the VVM size and the subsequent cloud-side verification workload grow linearly ($O(n)$) with ECU density. For a 100-ECU Zonal SDV, the Director Repository must verify 100 independent signatures per vehicle check-in, creating the computational and communication bottleneck that zk-ScalHard is designed to eliminate through recursive aggregation.
\begin{algorithm}[H]
\caption{Uptane OTA Update Request Lifecycle}
\label{alg:uptane_flow}
\begin{algorithmic}[1]
\REQUIRE Software update request for $n$ ECUs.
\ENSURE Fleet-wide authentication and target identification.

\STATE \textbf{Tier-0 (Edge):} Each secondary ECU generates a signed EVR. (See Algorithm \ref{alg:evr_gen} in Appendix).
\STATE \textbf{Tier-2 (Hub):} Primary ECU collects $n$ EVRs and bundles them into a signed VVM. (See Algorithm \ref{alg:vvm_gen} in Appendix).
\STATE \textbf{Tier-3 (Cloud):} Director Repository validates the VVM against the inventory database and determines the required software images for all $n$ ECUs. (See Algorithm \ref{alg:cloud_verif} in Appendix).
\end{algorithmic}
\end{algorithm}

\subsubsection{EVR Framework Structure}
\begin{figure}[t]
  \centering
  \includegraphics[width=\linewidth]{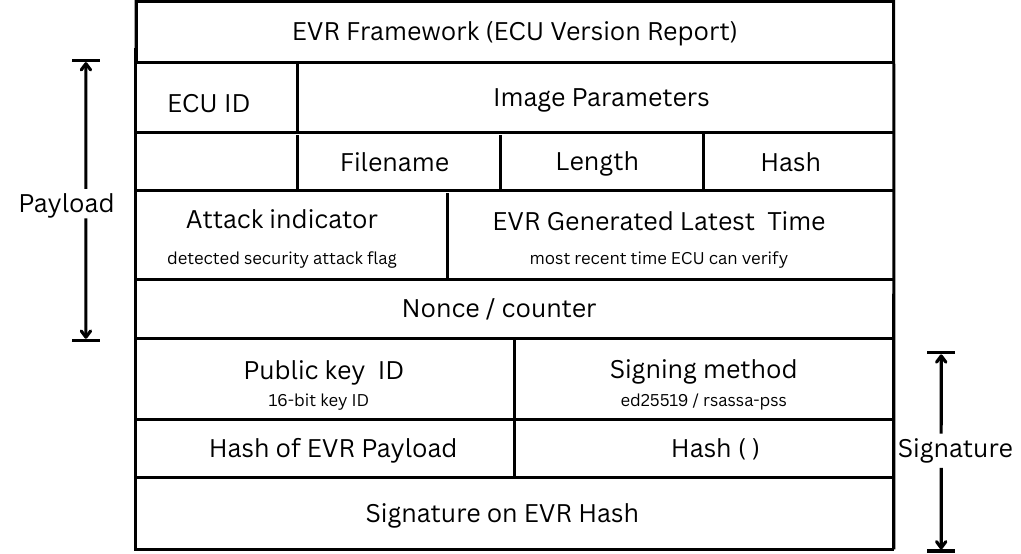}
  \caption{The individual ECU report (EVR)}
  \label{fig:evr1}
\end{figure}

The ECU Version Report (EVR) is the fundamental unit of attestation in Uptane. As shown in Fig. \ref{fig:evr1}, each EVR contains a physical hardware ID, firmware metadata, and a unique digital signature. While secure for a single node, the requirement that each ECU generate and transmit a full cryptographic signature imposes a significant computational burden at the edge. The comprehensive attestation sequence for edge-node metadata is formalized in Algorithm \ref{alg:evr_gen} (Appendix A.1). 

\subsubsection{VVM Framework Structure}
\begin{figure}[t]
  \centering
  \includegraphics[width=\linewidth]{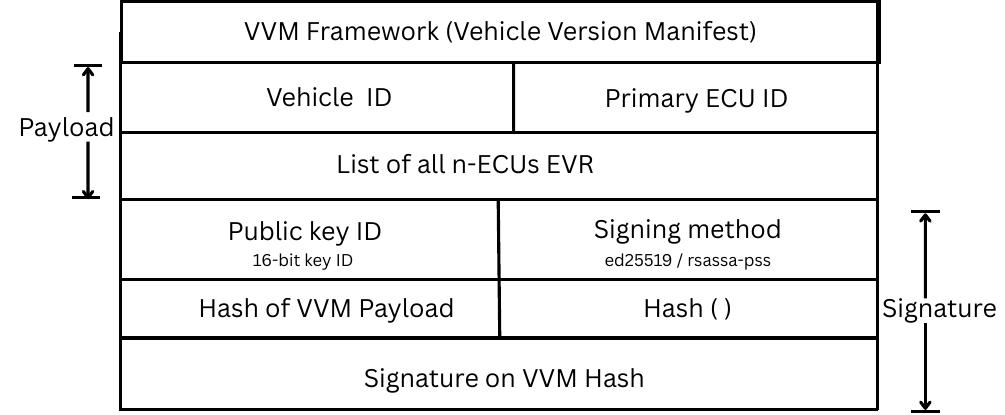}
  \caption{The aggregated vehicle manifest, illustrating the linear $O(n)$ growth of the VVM payload}
  \label{fig:vvm1}
\end{figure}

The Vehicle Version Manifest (VVM) is the top-level metadata object transmitted to the cloud. As illustrated in Fig. \ref{fig:vvm1}, the VVM is a linear concatenation of all $n$ EVRs within the vehicle. This structure is the primary source of the linear scalability crisis; since the VVM size is $O(n)$, the communication overhead and cloud-side verification time grow proportionally with ECU density. The VVM aggregation and signing logic is delineated in Algorithm \ref{alg:vvm_gen} (Appendix A.2). 

\subsection{ZKP Primitives}
Zero-Knowledge Proofs (ZKPs) allow a Prover to convince a Verifier of the validity of a statement without revealing the underlying secret (the witness). Our protocol utilizes a hybrid selection of ZKP primitives:
\begin{itemize}
\item Groth16: A pairing-based zk-SNARK known for its ultra-succinct proof size (256 bytes) and constant-time verification. We utilize Groth16 for Tier-1 (ZCU-to-HPC) communication where internal bus bandwidth is the primary constraint. It requires a one-time "Trusted Setup," which we decentralize via our Zonal MPC ceremony.

\item Plonky3 / FRI-based SNARKs: These are "transparent" protocols that do not require a trusted setup and offer Post-Quantum (PQ) resilience by relying on hash-based polynomial commitments (FRI). We utilize these at Tier-2 (HPC-to-Cloud) to provide long-term security against quantum adversaries and to enable Recursive Aggregation, allowing the vehicle to compress multiple zonal proofs into a single vehicle-level attestation.
\end{itemize}

\section{Adversary Model and Design Goals}
In this section, we formalize a robust adversarial model \begin{math}'A'\end{math} spanning the vehicle's physical silicon (Layer-0) to the OEM's OTA cloud infrastructure (Layer-3). Furthermore, we establish the core design requirements for achieving a secure, scalable, and privacy-preserving authentication protocol for OTA updates in Zonal SDV.

\subsection{Adversary Model}
We categorize the adversarial threat landscape into two primary vectors based on the Zonal SDV's architectural topology: vertical (Silicon-to-Cloud) and horizontal (Device-to-Network). This dual-perspective approach ensures a comprehensive analysis of both physical-layer vulnerabilities and logical network exploits.
\subsubsection{Vertical Threat Vector (Silicon-to-Cloud)}
\label{sec:verticalThreat}
\begin{itemize}
\item Physical Level ({\itshape The Silicon}): The adversary \begin{math}'A'\end{math} is assumed to have temporary physical access to the vehicle’s ECUs/ZCUs (e.g., during maintenance or at a service station). This enables ‘A’ to execute physical attacks, including voltage glitching and side-channel analysis (SCA), on persistent flash memory to extract digital secrets, such as stored private keys.
\item Infrastructure Level ({\itshape The Cloud}): \begin{math}'A'\end{math} may act as an external attacker targeting the centralized OEM’s OTA server or act as a malicious insider with administrative access privileges. The main goal of ‘A’ is to compromise the "Toxic Waste" (the secret parameters of the trusted setup) to enable the forgery of authentication proofs for the entire vehicle fleet.
\end{itemize}

\subsubsection{Horizontal Threat Vector (Node-to-Network)}
\label{sec:horizontalThreat}

\begin{itemize}
\item Node Level ({\itshape Central HPC}): The central HPC in a Zonal SDV serves as the primary logical attack surface, as it acts as the gateway between the internal zonal backbone and external network interfaces. We assume \begin{math}'A'\end{math} can exploit software vulnerabilities during the OTA update process to achieve root-level privilege escalation within the HPC’s execution environment. In this compromised state, the \begin{math}'A'\end{math} acts as a malicious aggregator, capable of intercepting internal zonal metadata or attempting to bypass the secure boot sequence to gain persistent remote control over the vehicle’s central computing functions.

\item Network Level ({\itshape IV/V2I Communication}): We adopt the Dolev-Yao adversary model to characterize threats across both in-vehicle (IV) and vehicle-to-infrastructure (V2I) channels. In this model, the adversary \begin{math}'A'\end{math} can sniff, intercept, and replay messages transmitted over the internal CAN-FD or Automotive Ethernet backbones. Furthermore, \begin{math}'A'\end{math} can manipulate V2I traffic exchanged via the Internet between the vehicle and the OEM cloud. By executing Man-in-the-Middle (MITM) and replay attacks, the adversary aims to capture valid zk-ScalHard proofs and reuse them to impersonate a legitimate ECU or ZCU, thereby compromising the overall integrity of the Zonal SDV.
\end{itemize}

\subsection{Design Goals}
To mitigate the aforementioned vertical and horizontal threat vectors, we establish the following four system-level security and performance requirements for the {\itshape zk-ScalHard} protocol.

\begin{itemize}
\item Goal 1. {\itshape Hierarchical Trust Promotion}: The vehicle authentication protocol must establish a multi-tier trust hierarchy that bridges the gap between resource-constrained edge ECUs (Layer-0) and high-performance OTA server (Layer-3). The objective is to align cryptographic intensity with heterogeneous hardware capabilities, ensuring that the overall system security is orchestrated from the silicon level up to the cloud without overburdening low-power edge nodes. 

\item Goal 2. {\itshape Asymptotic Scalability}: The protocol must achieve constant-size $O(1)$ complexity for both communication overhead and verification latency. This requirement ensures that the authentication tax remains invariant as the number of vehicle edge ECUs $(n)$ increases, effectively decoupling the security overhead from the Zonal SDV's architectural complexity.
\item Goal 3. {\itshape Decentralized Trust Initialization}: The protocol must facilitate the in-situ generation of unique ZKP proving $(pk)$ and verification $(vk)$ keys within each ZCU. By deriving the necessary secret randomness (i.e., 'toxic waste') locally via multi-party computation, the system must eliminate the systemic single point of failure inherent to centralized trusted setups. This ensures that a compromise of a global server cannot be leveraged to forge credentials for the entire vehicle fleet.
\item Goal 4. {\itshape Temporal Secret Isolation and Sovereignty}: The protocol must ensure that the unique secrets (witnesses) of vehicle nodes are ephemeral, generated on-the-fly, and resident in volatile memory only during the transient execution window of the ZKP generation. By eliminating the reliance on static private keys resident in persistent Flash memory, the system must minimize the temporal attack surface for physical extraction. This approach ensures 100\% vehicle data sovereignty, protecting the vehicle's physical identity from long-term side-channel analysis and memory-dump exploits.

\end{itemize}

\section{The zk-ScalHard Architecture - The "Trust Pyramid"}
\begin{figure}[t]
  \centering
  \includegraphics[width=\linewidth]{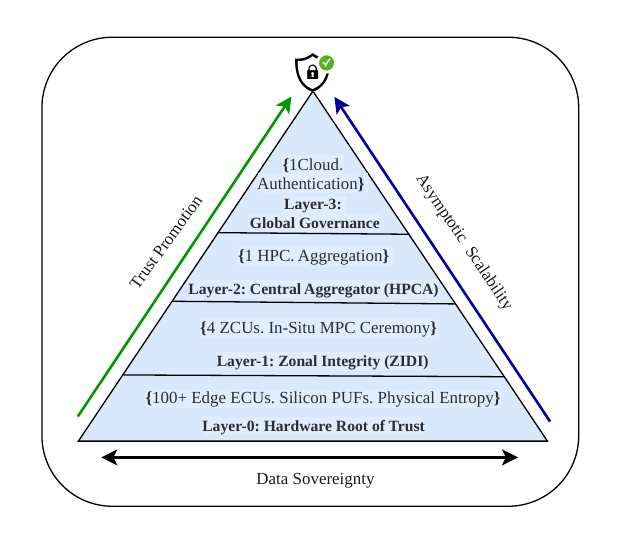}
  \caption{The zk-ScalHard Hierarchical Trust Pyramid}
  \label{fig:trust_pyramid_arc}
\end{figure}

\begin{figure}[t]
  \centering
  \includegraphics[width=\linewidth]{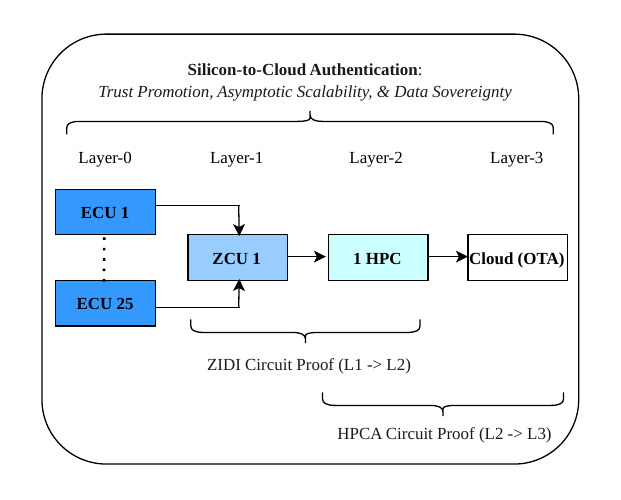}
  \caption{Vehicle-to-Cloud Authentication During Secure OTA Update
Lifecycle}
  \label{fig:v2i_auth_arc}
\end{figure}

We introduce {\itshape zk-ScalHard}, a novel multi-tier hierarchical trust pyramid architecture (Fig. \ref{fig:trust_pyramid_arc}) anchored in a ‘Silicon-to-Cloud’ security philosophy. This architecture is designed to support scalable vehicle-to-cloud authentication throughout the secure OTA update lifecycle in Zonal SDVs. As illustrated in the authentication flow (Fig. \ref{fig:v2i_auth_arc}), it orchestrates trust from the physical hardware (edge ECUs) to the global cloud (OTA server) based on three primary design principles: (i) \textbf{Trust promotion} through hierarchical attestation, (ii) \textbf{Asymptotic scalability} leveraging recursive proof aggregation, and (iii) \textbf{Data sovereignty} via hardware-rooted silicon PUFs. 

Upon initiating an OTA update request, the architecture escalates cryptographic strength across tiers: edge ECUs (Layer-0) first perform symmetric MAC-based authentication to their respective ZCUs. Subsequently, each ZCU (Layer-1) authenticates to the central HPC via a zk-SNARK-based ZIDI proof. Finally, the vehicle’s central HPC (Layer-2) aggregates these zonal proofs into a single, asymptotically compressed HPCA proof for verification by the global OTA server (Layer-3). The detailed operations of each layer are discussed below.

\subsection{Layer-0: The Hardware Root of Trust}
This layer anchors each edge ECU's identity to its unique Silicon PUF, replacing persistent digital keys. These physical secrets are never stored in non-volatile Flash memory; instead, they are reconstructed on-the-fly in response to an OTA update request. The PUF response serves a dual purpose: (i) it provides the entropy for symmetric MAC generation during local ZCU authentication, and (ii) it serves as the source of noisy entropy (i.e., ‘toxic waste’) for the in-situ Zonal MPC ceremony executed at Layer-1.

\subsection{Layer-1: The Zonal Integrity (ZIDI)}
In this tier, each ZCU operates as a ‘Zonal Judge’ responsible for validating the integrity of its local zone sub-ECU. To establish a decentralized root-of-trust, each ZCU executes an in-situ ‘Zonal MPC Ceremony’ (offline Phase 2, see Section 5) utilizing the noisy PUF entropy from its 25 local edge ECUs to generate a vehicle’s zonal-unique pair of proving $(pk)$ and verification $(vk)$ keys. This process ensures that the ‘toxic waste’ is consumed and destroyed locally within the zonal hardware. 

During the runtime OTA update cycle (online Phase 3, see Section 5), the ZCU verifies the 25 sub-ECU MACs and utilizes the ZIDI circuit to generate a succinct 809-byte proof. This proof is constructed using the ZCU’s own Silicon PUF as a ‘private witness’, which is reconstructed on-the-fly and immediately purged from volatile memory post-execution. The ZCU subsequently transmits only the proof to the central HPC, which utilizes the pre-distributed $vk$ to verify the zone’s integrity without ever accessing the raw silicon identities.

\subsection{Layer-2: Central Aggregation (HPCA)}
The Zonal SDV’s HPC serves as the vehicle’s central processing hub. It acts as a ‘transparent’ aggregator, consolidating the vehicle's global security state to achieve asymptotic $O(1)$ scalability. Upon receiving the four ZIDI proofs from the Layer-1 ZCUs- collectively representing the integrity of 100 edge ECUs- the HPC executes the HPCA circuit. 

This circuit performs recursive aggregation and mathematically validates the four zonal proofs while simultaneously integrating the HPC’s on-the-fly PUF-derived witness. This process culminates in a single, vehicle-level proof of constant size (805 bytes) that encapsulates the metadata integrity of the entire vehicle (100 ECUs, 4 ZCUs, and the HPC). The resulting succinct proof is then transmitted to the OTA cloud server for final vehicle-to-cloud authentication.

\subsection{Layer-3: Global Governance} In this final tier, the OEM’s OTA cloud server acts as the global verifier for the vehicle fleet. Upon receiving the constant-size proof from the vehicle’s HPC, the cloud retrieves the corresponding verification key $(vk)$ from its ‘Digital Twin’ database (originally generated during the offline Phase-2 Factory Ceremony, see Section 5), to validate the attestation. Because the vehicle transmits a single aggregated proof, the cloud performs only a single $O(1)$ verification. This ensures that the cloud-side computational tax remains invariant regardless of the internal vehicle's complexity (i.e., the number of ECUs and ZCUs).  

Ultimately, the OTA server only learns the vehicle's binary integrity state; the granular metadata on ECU versions and raw silicon PUF responses remains hidden. This architecture guarantees 100\% data sovereignty and full compliance with the GDPR mandates by ensuring that no personally identifiable information (PII) or sensitive configuration data ever leaves the vehicle’s secure environment.

\section{Technical Design and Implementation}
This section describes the formal technical design and implementation of the {\itshape zk-ScalHard protocol}. We first formalize the arithmetic logic for the {\itshape ZIDI} and {\itshape HPCA} circuits, which underpin our claims of decentralized trust initialization and asymptotic scalability. Furthermore, we discuss the {\itshape temporal secret isolation} mechanism that governs the lifecycle of the physical PUF witness. Finally, we describe the implementation specifics, including the arithmetic primitives and the simulation environment used to characterize the protocol’s performance.
\subsection{ZIDI Circuit: Zonal Integrity and Trust Initialization} 
The ZIDI circuit design is divided into three phases: Phase 1. Global arithmetic circuit design (offline/ compile-time), Phase 2. In-Situ trust initialization (offline/factory-enrollment), and Phase 3. Runtime authentication (online/ update-cycle). The detailed description of these phases is as follows. 
\subsubsection{Phase-1: Global arithmetic circuit design (offline/compile-time)} This phase focuses on the ZIDI arithmetic logic, implemented in Circom 2.1. The OEM engineers develop this circuit once (offline) and use it across all vehicles, serving as a universal security blueprint. While the blueprint is universal, unique zonal ZKP parameters are generated for each vehicle instance. To accommodate the Zonal SDV’s resource-constrained environment, we implemented a ‘tiered 5x5 hashing’ strategy to aggregate the 25 sub-ECU MACs into five sub-groups using the Poseidon hash function. This tiered approach overcomes the 16-input limit of the Poseidon library primitives. By leveraging optimized permutation constants, the circuit maintains high integrity within a hierarchical structure.

Compiling the ZIDI circuit into a Rank-1 Constraint System (R1CS) yields 2,163 non-linear constraints. This succinct constraint count makes it feasible to execute the circuit on 32-bit automotive microcontrollers (ECUs/ZCUs). The algorithmic steps for Phase 1 are detailed in Algorithm \ref{alg:zidi_logic}.

\subsubsection{Phase-2: In-Situ trust initialization (offline/factory enrollment)}
In this phase, the OEM executes the universal ZIDI circuit’s R1CS within each vehicle zone to initiate the In-Situ Zonal MPC Ceremony and generate unique ZKP parameters $(pk, vk)$. This one-time factory enrollment (offline) establishes a unique cryptographic trust relationship within the vehicle's zonal hardware. The process begins with the ZCU collecting noisy silicon PUF responses from 25 local edge ECUs as physical entropy to derive the secret randomness \begin{math}T\end{math} (called ‘toxic waste’ ). To accommodate the tiered Poseidon logic defined in Phase 1, we utilize a ‘Powers of Tau’ setup with 2\textsuperscript{14} (16,384) constraint capacity. 

We utilize the Groth16 setup protocol, taking R1CS and \begin{math}T\end{math} as inputs to generate a zonal-unique pair of ZKP parameters $(pk, vk)$. The proving key $(pk)$ is provisioned to the ZCUs’ secure Flash memory for use in Phase 3 proof generation, while the verification key $(vk)$ is stored within the vehicle’s central HPC for the subsequent verification phase. Finally, the master entropy (\begin{math}T\end{math}) is immediately purged from the volatile memory, ensuring that even the OEM or OTA server cannot forge proofs for the vehicle during its OTA update lifecycle. The algorithmic steps for Phase 2 are detailed in Algorithm \ref{alg:mpc_ceremony}.

\begin{algorithm}[t]
\caption{ZIDI Circuit Arithmetic Constraints ($\mathcal{C}_{ZIDI}$)}
\label{alg:zidi_logic}
\begin{algorithmic}[1]
\REQUIRE \textbf{Private Signals (Witness $W$):} 
\STATE $W_{puf} \in \mathbb{F}_p$ \COMMENT{Silicon PUF stable bitstring}
\STATE $W_{macs}[25] \in \mathbb{F}_p^{25}$ \COMMENT{Array of 25 Edge-ECU MACs}
\REQUIRE \textbf{Public Signals (Statement $X$):}
\STATE $ID_{zcu} \in \mathbb{F}_p$ \COMMENT{Registered Zonal Identity}
\STATE $H_{agg} \in \mathbb{F}_p$ \COMMENT{Expected Zonal Integrity Root}
\STATE $N \in \mathbb{F}_p$ \COMMENT{Freshness Nonce from HPC}

\STATE \textbf{Constraint 1: Identity Anchoring}
\STATE $h_{puf} \leftarrow \text{Poseidon}_1(W_{puf})$
\STATE \textbf{assert} $h_{puf} \equiv ID_{zcu}$ \COMMENT{Ensures hardware-software binding}

\STATE \textbf{Constraint 2: Tiered Integrity Aggregation}
\FOR{$i = 0$ \TO $4$}
    \STATE \COMMENT{Level 1: Hash 5 groups of 5 ECUs}
    \STATE $h_{sub, i} \leftarrow \text{Poseidon}_5(W_{macs}[5i \dots 5i+4])$
\ENDFOR
\STATE \COMMENT{Level 2: Compute Zonal Root from sub-hashes}
\STATE $h_{root} \leftarrow \text{Poseidon}_5(h_{sub, 0}, h_{sub, 1}, h_{sub, 2}, h_{sub, 3}, h_{sub, 4})$
\STATE \textbf{assert} $h_{root} \equiv H_{agg}$ \COMMENT{Validates 25 ECU states in $O(1)$}

\STATE \textbf{Constraint 3: Temporal Binding}
\STATE \textbf{dummy} $\leftarrow N \times N$ \COMMENT{Forces Nonce into R1CS to prevent replay}

\RETURN $\pi_{zonal}$ \COMMENT{Succinct Zonal Proof}
\end{algorithmic}
\end{algorithm}

\subsubsection{Phase-3: Runtime authentication (online/ update-cycle)}
This phase is an online operation in which each ZCU performs a 2-way, stateless handshake with the central HPC during an OTA update. The process begins with the HPC (verifier) issuing a ‘Nonce’—a unique 256-bit random challenge to prevent replay attacks. Upon receiving the ‘Nonce’, the ZCU initiates proof generation by executing the ZIDI prover function with four inputs: (i) its stable silicon PUF bits, reconstructed on-the-fly to serve as a dynamic private witness, (ii) 25 local ECU MACs (from Phase 2), (iii) its unique proving key $(pk)$, and (iv) the ‘Nonce’. This process results in a succinct 809-byte zk-proof ($\pi$ ). The ZCU transmits $\pi$  to the HPC, which verifies the proof using the corresponding verification key $(vk)$. Crucially, the HPC achieves constant verification complexity $O(1)$, as the mathematical operations remain invariant regardless of the number of ECUs (n=25). While the total ZCU proving latency is \begin{math} \approx 4.2s \end{math}, the verification is near-instant. 

Throughout this process, raw PUF bits are never transmitted across the vehicle network, ensuring 100\% vehicle data sovereignty. Finally, the ZCU purges the PUF witness from RAM, achieving temporal isolation. These Phase-3 steps are formalized in Algorithm \ref{alg:handshake}.

\begin{algorithm}[t]
\caption{In-Situ Zonal MPC Trust Initialization}
\label{alg:mpc_ceremony}
\begin{algorithmic}[1]
\REQUIRE \textbf{Entity:} Zonal Controller ($ZCU_j$), Child ECUs $\{E_1 \dots E_{25}\}$
\REQUIRE \textbf{Input:} Universal R1CS blueprint $\mathcal{C}_{ZIDI}$
\STATE \textbf{Step 1: Physical Entropy Collection}
\FOR{each $E_i \in \{E_1 \dots E_{25}\}$}
    \STATE $S_i \leftarrow \text{Trigger\_PUF}(E_i)$ \COMMENT{Extract physical silicon noise}
\ENDFOR
\STATE $\tau \leftarrow \text{Poseidon}(\sum S_i)$ \COMMENT{Derive local 'Toxic Waste'}
\STATE \textbf{Step 2: Parameter Generation (Groth16)}
\STATE $\{pk_j, vk_j\} \leftarrow \text{Groth16.Setup}(\mathcal{C}_{ZIDI}, \tau)$
\STATE \textbf{Step 3: Secure Provisioning}
\STATE $\text{Write}(pk_j) \to \text{ZCU}_j.\text{SecureFlash}$
\STATE $\text{Write}(vk_j) \to \text{HPC}.\text{TrustStore}$
\STATE \textbf{Step 4: Anti-Forensic Purge}
\STATE $\text{Erase}(\tau)$ \COMMENT{Master secret destroyed locally}
\ENSURE Unique Zonal Parameters $\{pk_j, vk_j\}$
\end{algorithmic}
\end{algorithm}

\begin{algorithm}[t]
\caption{Tier-1: Zonal Integrity Handshake}
\label{alg:handshake}
\begin{algorithmic}[1]
\REQUIRE \textbf{Prover:} $ZCU_j$ with $pk_j$; \textbf{Verifier:} $HPC$ with $vk_j$
\STATE \textbf{HPC $\to$ ZCU:} Request Update + $Nonce$ ($N$)
\STATE \textbf{ZCU Operation:}
\STATE $W_{puf} \leftarrow \text{Reconstruct\_PUF}(ZCU_j)$
\STATE $W_{macs} \leftarrow \text{Collect\_ECU\_MACs}(E_{1 \dots 25})$
\STATE $\pi_{zonal} \leftarrow \text{Prove}(pk_j, [W_{puf}, W_{macs}], [ID, H_{state}, N])$
\STATE \textbf{ZCU $\to$ HPC:} $\pi_{zonal}$ \COMMENT{Succinct 809-byte proof}
\STATE \textbf{HPC Operation:}
\IF{$\text{Verify}(vk_j, \pi_{zonal}, N) \equiv 1$}
    \STATE \textbf{return} \texttt{ZONE\_AUTHENTICATED}
\ELSE
    \STATE \textbf{return} \texttt{ABORT\_SESSION}
\ENDIF
\STATE \textbf{ZCU:} $\text{Purge}(W_{puf})$ from RAM \COMMENT{Temporal Isolation}
\end{algorithmic}
\end{algorithm}

\subsection{HPCA Circuit: Recursive Proofs Aggregation} 
The central HPC acts as a ‘Recursive Aggregator’, executing the HPCA circuit to aggregate the proofs from the four Layer-1 ZCUs. The circuit consumes three primary inputs: (i) the four ZIDI zonal proofs, (ii) the HPC’s stable PUF response, reconstructed on-the-fly as a private witness, and (iii) the proving key $(pk)$. 

The HPCA circuit is extremely efficient, consisting of only 518 non-linear R1CS constraints. The HPCA prover function mathematically folds these zonal states into a single, vehicle-to-cloud proof (\begin{math} \approx 805 bytes\end{math}). This succinct proof encapsulates the vehicle's global integrity state (covering 100 ECUs, 4 ZCUs, and the HPC) and achieves asymptotic $O(1)$ scalability. Finally, the HPC transmits this proof to the OTA cloud server, enabling the cloud to verify the entire vehicle state in constant time using the corresponding verification key $(vk)$. These Phase-3 steps are formalized in Algorithm \ref{alg:hpca}.

\begin{algorithm}[H]
\caption{Tier-2: HPCA Recursive Aggregation}
\label{alg:hpca}
\begin{algorithmic}[1]
\REQUIRE \textbf{Prover:} $HPC$; \textbf{Verifier:} $Cloud$ 
\REQUIRE \textbf{Input:} 4 Zonal Proofs $\{\pi_{zonal, 1 \dots 4}\}$
\REQUIRE \textbf{Circuit:} $\mathcal{C}_{HPCA}$ with Proving Key $pk_{hpc}$
\STATE \textbf{Step 1: Inner Proof Verification}
\FOR{$j = 1$ \TO $4$}
    \STATE \textbf{assert} $\text{Verify}_{ZIDI}(vk_j, \pi_{zonal, j}) \equiv 1$
\ENDFOR
\STATE \textbf{Step 2: Identity \& State Folding}
\STATE $W_{hpc} \leftarrow \text{Reconstruct\_PUF}(HPC)$
\STATE $H_{global} \leftarrow \text{Poseidon}(Root_1, \dots, Root_4)$
\STATE \textbf{Step 3: Global Proof Generation}
\STATE $\pi_{veh} \leftarrow \text{Prove}(pk_{hpc}, [\{\pi_{zonal}\}, W_{hpc}], [ID_{veh}, H_{global}])$
\STATE \textbf{HPC $\to$ Cloud:} $\pi_{veh}$ \COMMENT{Constant-size 805-byte proof}
\ENSURE Asymptotic $O(1)$ Vehicle-to-Cloud Attestation
\end{algorithmic}
\end{algorithm}

\subsection{Temporal Secret Isolation Mechanism}
The {\itshape zk-ScalHard} protocol utilizes a hardware-rooted, device-unique silicon PUF as a ‘stateless witness’, in contrast to legacy systems that rely on persistent digital keys stored in non-volatile Flash memory. During the vehicle’s idle state (no active OTA requests), no secrets reside in volatile or non-volatile memory, effectively maintaining a ‘zero-secret attack surface’. The device’s physical PUF response is triggered only upon a valid OTA update request; the secret is reconstructed on the fly in volatile memory, which takes $357ms$. Subsequently, the ZKP prover (ZIDI or HPCA) consumes this witness to generate the proof over 3,885ms. Immediately upon proof completion, the witness is ‘securely purged’ from the volatile memory. Consequently, the secret identity is present in the system for only \begin{math} \approx 4.2s \end{math} during the OTA update lifecycle. 

We demonstrated this (see Section 7, Fig. \ref{fig:temporal}) and compared it with legacy secret exposure over 24 hours. Our protocol achieves a 99.9\% reduction in the temporal attack surface.

\subsection{Implementation and Benchmarking Environment} 
The {\itshape zk-ScalHard} protocol’s ZIDI and HPCA circuits are implemented using the Circom 2.1.6 hardware description language and the Poseidon-2 hash primitive to generate universal R1CS constraints. We utilize the Groth16 ZKP proving system and the SnarkJS framework to generate the proving $(pk)$ and verification $(vk)$ parameters, ensuring a constant succinct proof size \begin{math} \approx 805 bytes\end{math} and constant verification complexity. 

To evaluate the protocol, we conducted Software-in-Loop (SIL) benchmarks within a virtualized Ubuntu 22.04 LTS environment via WSL2.  The host system specifications included an 11th Gen Intel Core i5-1135G7 CPU @ 2.40GHz, 8GB of RAM, and Windows 10.  For the legacy baseline, we utilized the Uptane Python reference implementation to quantify the linear $O(n)$ metadata overhead and verification latencies. Finally, to facilitate reproducibility and support open science, the official implementation of zk-ScalHard and benchmarking scripts are hosted in a 
public GitHub repository: \url{https://anonymous.4open.science/r/zk-ScalHard-8C52/README.md}

\section{Security and Resilience Analysis}
We evaluate the security posture of zk-ScalHard against the adversary model $\mathcal{A}$ defined in Section 3. Table \ref{tab:security_matrix} provides a comprehensive mapping of identified threat vectors to our protocol's specific mitigation mechanisms. This matrix ensures that the hierarchical trust model provides end-to-end coverage, from physical silicon integrity to global cloud-level verification.

\begin{table}[h]
\centering
\caption{Security Matrix: Mapping Threats to zk-ScalHard Mechanisms}
\label{tab:security_matrix}
\resizebox{\columnwidth}{!}{
\begin{tabular}{|l|l|l|}
\hline
\textbf{Threat Vector (Sec 3)} & \textbf{zk-ScalHard Mechanism} & \textbf{Security Property} \\ \hline
Physical Key Extraction & Silicon PUF + Temporal Isolation & Physical Unforgeability \\ \hline
Network MITM / Replay & Nonce-Bound R1CS Constraints & Strong Freshness \\ \hline
Malicious HPC Aggregator & Recursive Proof Verification & Integrity Isolation \\ \hline
Infrastructure Leakage & In-Situ Zonal MPC Ceremony & Decentralized Trust \\ \hline
Quantum Forgery & FRI-based PQ-ZKP (Tier-2) & PQ-Resilience \\ \hline
\end{tabular}
}
\end{table}

\subsection{Formal Security Properties}
In this section, we formalize the security of the zk-ScalHard protocol through the following theorems.

\newtheorem{theorem}{Theorem}

\begin{theorem}[Completeness]
If the Prover (ZCU/HPC) and the Verifier (HPC/Cloud) are honest and follow the zk-ScalHard protocol, the verification of the aggregated proof $\pi_{veh}$ will always succeed with probability 1.

The completeness of zk-ScalHard is derived from the deterministic nature of the ZIDI and HPCA arithmetic circuits. Given valid PUF witnesses and correct ECU MACs, the Poseidon-based constraints will always satisfy the R1CS requirements, resulting in a valid proof $\pi$.
\end{theorem}

\begin{theorem}[Soundness]
For any probabilistic polynomial-time (PPT) adversary $\mathcal{A}$, the probability that $\mathcal{A}$ can generate a valid proof $\pi$ without knowledge of a valid Silicon PUF witness or a valid set of ECU MACs is negligible.

The soundness relies on the collision-resistance of the Poseidon hash function and the discrete-logarithm hardness of the Groth16/Plonky3 schemes. An adversary cannot forge $\pi$ without the physical PUF response, as the probability of a witness collision is negligible.
\end{theorem}

\begin{theorem}[Zero-Knowledge]
The zk-ScalHard protocol is zero-knowledge; the Verifier gains no information about the vehicle's internal configuration (ECU versions) or raw Silicon PUF responses, other than the validity of the integrity state.

The protocol achieves computational zero-knowledge by ensuring that the proof $\pi$ is a succinct commitment that does not leak the underlying witness $\mathcal{W}$. The verifier only learns the binary validity of the state, preserving 100\% vehicle configuration privacy.
\end{theorem}

\subsection{Resilience to Physical Key Extraction}
As quantified in Fig. \ref{fig:temporal}, we achieve temporal isolation. By ensuring the PUF witness exists only during the \begin{math} \approx 4.2s \end{math} proving window, we render physical probing attacks statistically infeasible.

\subsection{Resistance to Network-Layer Attacks}
Nonce-binding within the ZIDI circuit ensures that each proof is cryptographically tied to a unique session challenge, effectively neutralizing Man-in-the-Middle (MitM) and replay attacks.

\subsection{Integrity under Aggregator Compromise}
Even if the central HPC is logically compromised, it remains incapable of forging zonal proofs. The ZCUs maintain physical isolation of the PUF witnesses, ensuring that the 'malicious aggregator' cannot subvert the vehicle's global integrity state.

\subsection{Quantum-Blast-Radius Control}
By decentralizing the trust ceremony via Zonal MPC, we isolate the impact of a potential cloud-level parameter compromise. Each vehicle maintains a unique, in-situ initialized root-of-trust.

\section{Performance Evaluation}
In this section, we evaluate the performance of the {\itshape zk-ScalHard} protocol through a comparative analysis with the industry-standard Uptane framework. We utilize three primary metrics to characterize the system’s performance: (i) asymptotic communication complexity, (ii) asymptotic verification complexity, and (iii) temporal attack surface reduction. As detailed in Section 5.4, all benchmarks were conducted within a ‘SIL’ simulation environment. To test the architecture's scalability limits, we used a high-density configuration of 100 ECUs in the Zonal SDV model. 

The results empirically validate our claims of constant $O(1)$  scalability, hardware-rooted physical-layer resilience, and 100\% data sovereignty throughout the Zonal SDV’s OTA update lifecycle.

\subsection{Asymptotic Communication Complexity} 
\begin{figure}[t]
  \centering
  \includegraphics[width=\linewidth]{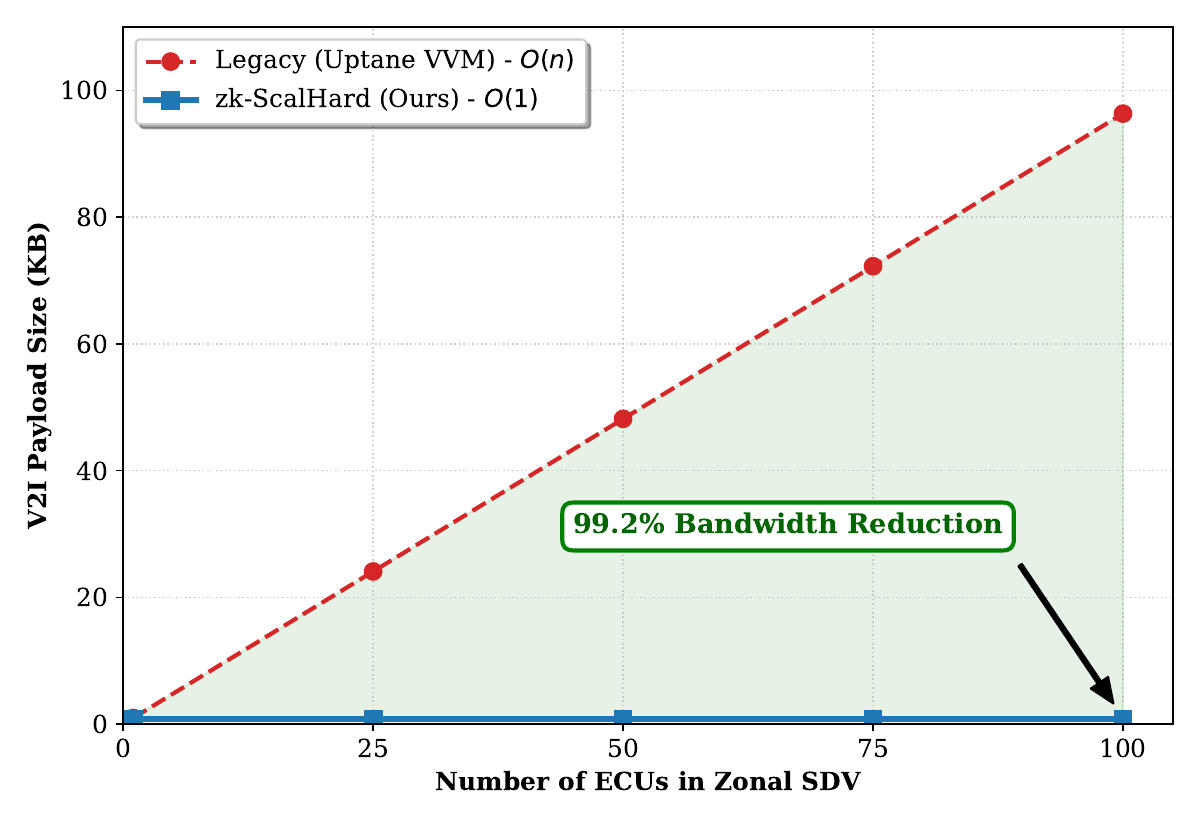}
  \caption{Asymptotic Communication Complexity}
  \label{fig:communication}
\end{figure}

Figure \ref{fig:communication} illustrates the communication-overhead advantages of {\itshape zk-ScalHard} within the Zonal SDV environment. Our benchmarks show that the V2I payload for the industry-standard Uptane framework grows linearly, $O(n)$, because the vehicle version manifest (VVM) must bundle individual reports and digital signatures for every ECU. For a 100-ECU configuration, the Uptane VVM reaches $98.6KB$. In comparison, {\itshape zk-ScalHard} achieves a constant-size $O(1)$ vehicle payload by utilizing a succinct HPCA proof of only $809B$. This represents a 99.2\% reduction in communication overhead. Consequently, {\itshape zk-ScalHard} ensures that network overhead remains invariant with respect to vehicle complexity, maintaining bus stability even as ECU density exceeds 100 nodes.

\subsection{Asymptotic Verification Complexity} 
\begin{figure}[t]
  \centering
  \includegraphics[width=\linewidth]{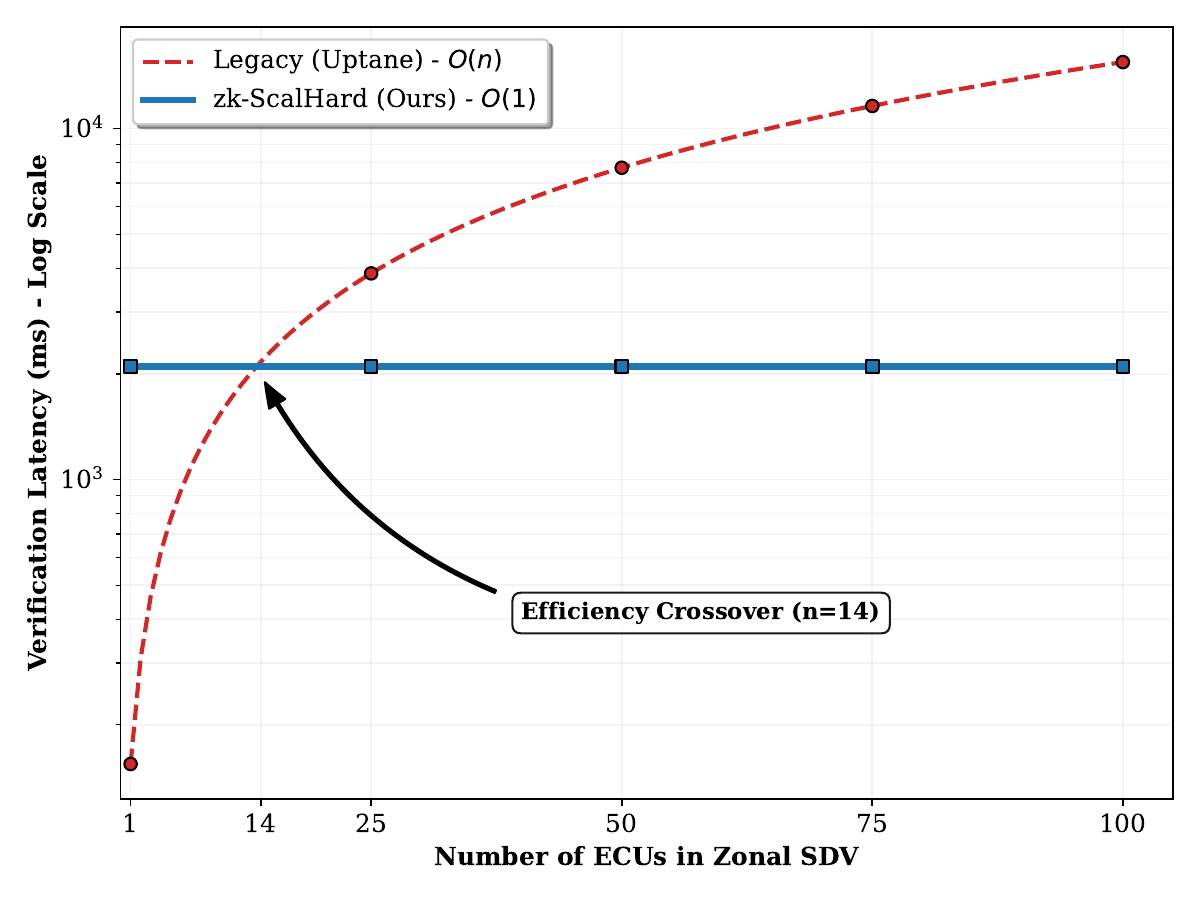}
  \caption{Verification Complexity: $O(n)$ vs. $O(1)$ Scalability}
  \label{fig:verification}
\end{figure}

We evaluate the computational scalability of the vehicle-cloud verification process, with results depicted in Fig. \ref{fig:verification}. Our results show that the Uptane framework suffers from a linear $O(n)$ verification bottleneck, requiring $15.47s$ to validate the 100 individual ECU signatures within a single VVM. This linear dependency makes legacy attestation unsustainable for high-density vehicle fleets.  In contrast, {\itshape zk-ScalHard} achieves constant verification complexity, $O(1)$. By using the HPCA circuit at HPC Layer-2 to aggregate zonal proofs, the OTA cloud server only performs a single mathematical verification.

As illustrated in Fig. \ref{fig:verification}, {\itshape zk-ScalHard} achieves an efficiency crossover at $n=14$ ECUs, outperforming legacy standards across all modern Zonal SDV configurations. While our prototype exhibits a constant simulation latency of $2.1s$ due to the high-level Node.js runtime and virtualized WSL environment, the mathematical complexity remains invariant. Specifically, at a scale of 100 ECUs, the comparison between {\itshape zk-ScalHard} $(2.1s)$ and Uptane $(15.47s)$ demonstrates a 7.3x computational speedup. This result confirms that {\itshape zk-ScalHard} is computationally superior to Uptane even within a non-optimized simulation environment. Furthermore, projecting these results onto production-grade ‘hardware security modules (HSMs)’, where verification is optimized to  \begin{math} \approx 5ms \end{math} (see Section 2), {\itshape zk-ScalHard} enables near-instantaneous fleet-wide attestation with zero computational growth.

\begin{figure}[ht]
  \centering
  \includegraphics[width=\linewidth]{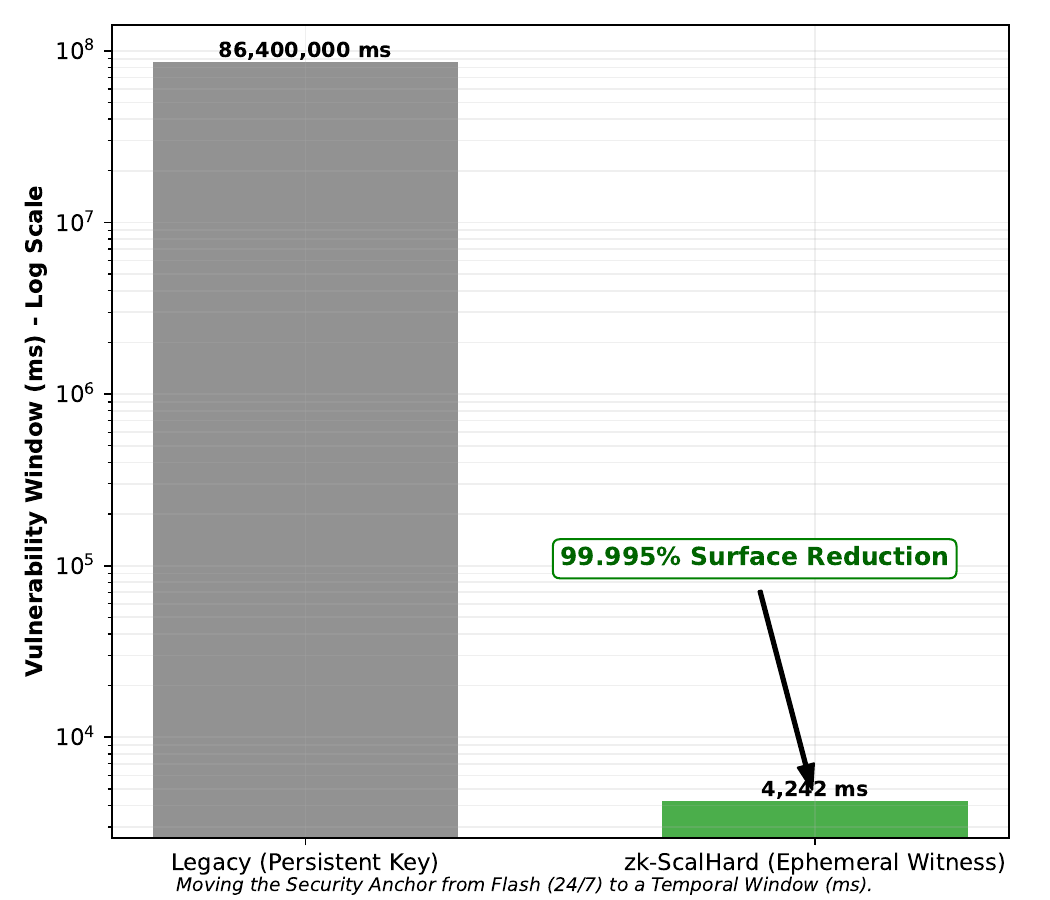}
  \caption{Temporal Attack Surface: Persistent vs. Ephemeral Secrets}
  \label{fig:temporal}
\end{figure}

\subsection{Temporal Attack Surface Reduction} 
Figure \ref{fig:temporal} quantifies the reduction in the ‘temporal vulnerability window’ by integrating Silicon PUFs as ephemeral witnesses, replacing persistent digital secrets. This architectural shift effectively neutralizes physical-layer threats. In legacy frameworks like Uptane, cryptographic keys are stored in non-volatile memory, resulting in a 24/7 exposure window $(8.64x10\textsuperscript{7}ms$ per day). Conversely, {\itshape zk-ScalHard’s} physical secrets are reconstructed on-the-fly and exist in volatile memory only during the ZK-proving window (\begin{math} \approx 4,242ms \end{math}). Our evaluation demonstrates a 99.99\% reduction in the temporal attack surface.  This ‘temporal isolation’ renders physical key extraction—such as side-channel analysis (SCA) or hardware probing—statistically infeasible, as the secret witness is securely purged before an adversary can successfully exploit it.

\section{Discussion and Compliance} 
zk-ScalHard is designed to align with the evolving automotive regulatory landscape. By achieving 100\% vehicle-level data sovereignty, the protocol satisfies the GDPR 'Data Minimization' principle, as sensitive metadata never leaves the vehicle's secure enclave. Furthermore, the temporal isolation of secrets addresses the 'Security-by-Design' mandates of the EU Cyber Resilience Act (CRA) and UNR 155/156, providing a robust defence against the full lifecycle of physical and remote threats in Zonal SDVs.

\section{Related Work}
The existing literature on SDVs has been extensively studied, and we categorise it into four domains: industry standards, privacy-preserving credentials, post-quantum cryptography, and ZKP-based frameworks.  While these domains provide robust solutions in isolation, the transition to Zonal SDV architecture (ZoA) demands a “hybrid model with collision of constraints” - high ECU density, limited \& high bandwidth (i.e., CAN-FD and high-speed Ethernet), and data privacy \& sovereignty needs. In this section, we evaluate these domain categories to highlight the trilemma gap -“Scalability, Privacy, and Resilience” that zk-ScalHard bridges.

\textbf{Industry Standards (Uptane, AUTOSAR):}
The current secure OTA updates industry standards are Uptane \cite{5Ref45,6Ref46,7Ref47,8Ref48,9Ref49,10Ref50} and AUTOSAR \cite{11Ref52,12Ref53,13Ref54}. Since both standards are built primarily using PKI infrastructure, these are robust frameworks. While Uptane provides secure OTA updates by solving the “Separation of Trust” problem, it is the current industry ‘gold standard.’ On the other hand, these standards' reliance on PKI-based linear signature accumulation leads to the ‘Metadata Blast’ we quantified in Section 1.2. This does not solve the $O(n)$ bandwidth issue; for a 100-ECU scale, the 98.6 KB payload becomes a network liability. Our zk-ScalHard protocol replaces the linear signature set with ‘Recursive Aggregation’ and is the first to achieve $O(1)$ communication complexity, a feature that current Uptane \cite{9Ref49} and AUTOSAR \cite{12Ref53} frameworks lack.

\textbf{Privacy-Preserving Credentials (Idemix, U-Prove):}
Existing privacy-preserving credential solutions like Idemix \cite{14Ref55,15Ref56,16Ref57} and U-Prove \cite{17Ref60} provide strong identity privacy using a ‘Zero-Knowledge’ approach. These are built for high-power platforms such as Hyperledger \cite{16Ref57}, cloud servers, and XML-heavy web protocols. Further, these are non-aggregable and computationally too heavy RSA math for resource-constrained 32-bit ZCUs. Where zk-ScalHard resolves this gap by utilizing ZKP-friendly Poseidon hashes \cite{18Ref24,19Ref25} and recursive SNARKs to aggregate 100+ ECUs' identities into a single succinct proof, allowing privacy-preserving attestation that fits within the limits of SDV’s control units.

\textbf{Post-Quantum Signatures:}
Recent efforts to NIST-standardize PQC like Dilithium and Falcon \cite{20Ref42} to defend against Shor’s algorithm and to integrate PQC signatures into V2X \cite{21Ref40}. These address the quantum threats but ignore the Physical Layer Constraints of the vehicle bus. That is, a single Dilithium signature is \begin{math} \approx 22.5 KB \end{math}; in Zonal SDV, with 100 ECU signatures, the result is 375 KB. This leads to high-density zonal attestation and on the vehicle’s internal CAN-FD network brick. zk-ScalHard achieves PQ-resilience using FRI-based ZKPs (Plonky3) while maintaining a significantly smaller, constant 805-byte footprint compared to linear PCQ signatures.

\textbf{ZKP Frameworks:}
Available ZKP frameworks \cite{22Ref35,23Ref61,24Ref62,25Ref63,26Ref64} are general-purpose and show the maturity of ZKP libraries. While most of these consider the underlying system to be ‘Flat Architecture’ and ignore the temporal attack surface. That is, they lack the ‘Hybrid Hierarchical Orchestration’ required for Zonal SDVs (ECU → ZCU → HPC → Cloud). Further, many existing ZKP-based solutions \cite{27Ref65,28Ref66,29Ref67,30Ref68,31Ref69} assume a single-tier prover-verifier model, and they don’t understand the ‘Trust Promotion’ requirement in the secure OTA updates for Zonal SDVs. Hence, these fail in Zonal SDV architecture. Our proposed zk-ScalHard solution is the first to propose a ‘Hybrid Hierarchical Trust Model’ that matches cryptographic intensity to specific hardware tiers (Symmetric → Succinct → Recursive). In addition, zk-ScalHard is the first to integrate PUFs as ‘Dynamic ZKP Witnesses’ in a Hierarchical Trust Pyramid, reducing the attack surface window from 24/7 to $4.2s$.



\section{Conclusion and Future Work}
We presented zk-ScalHard, a hardware-rooted ZKP protocol that achieves constant-size $O(1)$ scalability for SDV authentication. Our evaluation demonstrates a 99.2\% reduction in bandwidth and a 99.9\% reduction in the physical attack surface. Future work will focus on Hardware-in-the-Loop (HiL) validation within a specialized automotive security laboratory to characterize the protocol on production-grade vehicle network processors like NXP S32G.

\section*{Acknowledgment}
This work was supported by the MSCA Seal of Excellence (SoE) Fellowship at Inria Lille, France, funded by the University of Lille. The authors are grateful to the SERENDIPITY Team (Inria) and the SPRITZ Security \& Privacy Research Group (University of Padua) for the collaborative research environment. Shrikant Tangade acknowledges the support of the Anuvartik Mirji Bharatesh Institute of Technology (VTU-affiliated), Belagavi, India, for his sabbatical and the establishment of the autoMoTIVe-X Lab. We also thank the IEEE VTS Bangalore Chapter for its leadership in the Automotive Cybersecurity CoE initiative.




%

\bibliographystyle{IEEEtran}
\bibliography{references}








\appendices
\section{Uptane Framework}

\subsection{ECU Version Report (EVR) Generation}
Algorithm \ref{alg:evr_gen} delineates the attestation process at the hardware edge. Each secondary ECU is responsible for constructing a self-contained report that binds its current software state to its cryptographic identity. The requirement that each edge node compute an independent digital signature ($Sign$) contributes to the overall computational overhead at Tier-0.
\begin{algorithm}[H]
\caption{Tier-0: Secondary ECU Metadata Generation}
\label{alg:evr_gen}
\begin{algorithmic}[1]
\REQUIRE $ID_{ecu}, V_{img}, L_{img}, H_{img}, PR_{ecu}, PID_{ecu}$
\ENSURE Signed ECU Version Report ($EVR$).

\STATE \textbf{Init:} $ID \gets \{ID_{ecu}, PID_{ecu}\}$, $IMG \gets \{V_{img}, L_{img}, H_{img}\}$
\STATE \textbf{Check:} $FLAG \gets \text{Attack\_Detection}()$
\STATE \textbf{Time:} $T_{evr} \gets \text{EVR\_Time}()$
\STATE \textbf{Nonce:} $N \gets \text{Nonce}()$
\STATE \textbf{Payload:} $PL \gets \{ID_{ecu}, IMG, N, FLAG, T_{evr}\}$
\STATE \textbf{Integrity:} $H_{evr} \gets \text{Hash}(PL)$
\STATE \textbf{Auth:} $DSIG \gets \text{Sign}(PR_{ecu}, H_{evr})$
\STATE \textbf{Attr. Block:} $Sig_{attr} \gets \{PID_{ecu}, \text{Meth}, H_{evr}, \text{Func}, DSIG\}$
\STATE \textbf{Assembly:} $EVR \gets \{Sig_{attr}, PL\}$
\STATE \textbf{Transmit:} Send $EVR$ to Primary ECU.
\end{algorithmic}
\end{algorithm}

\subsection{Vehicle Version Manifest (VVM) Generation}
Algorithm \ref{alg:vvm_gen} formalizes the aggregation logic at the vehicle's central hub. The Primary ECU (HPC) passively bundles all $n$ incoming reports. This linear concatenation results in a VVM payload size that scales at $O(n)$, directly leading to the communication bottlenecks observed in legacy high-density SDV configurations.


\begin{algorithm}[H]
\caption{Tier-2: Primary ECU Metadata Aggregation}
\label{alg:vvm_gen}
\begin{algorithmic}[1]
\REQUIRE $VIN, ID_{pecu}, \{EVR_1, \dots, EVR_n\}, PR_{pecu}$
\ENSURE Signed Vehicle Version Manifest ($VVM$).

\STATE \textbf{Aggreg.:} $SET_{evr} \gets \{EVR_1, \dots, EVR_n\}$
\STATE \textbf{Self-Attest:} $EVR_{pecu} \gets \text{GenPrimaryReport}()$
\STATE \textbf{Payload:} $PL \gets \{VIN, ID_{pecu}, SET_{evr}, EVR_{pecu}\}$
\STATE \textbf{Integrity:} $H_{pvvm} \gets \text{Hash}(PL)$
\STATE \textbf{Auth:} $DSIG \gets \text{Sign}(PR_{pecu}, H_{pvvm})$
\STATE \textbf{Attr. Block:} $SIG_{attr} \gets \{ID_{pecu}, \text{Meth}, H_{pvvm}, \text{Func}, DSIG\}$
\STATE \textbf{Assembly:} $VVM \gets \{PL, SIG_{attr}\}$
\STATE \textbf{Transmit:} Send $VVM$ to Director Repository.
\end{algorithmic}
\end{algorithm}


\subsection{Director Repository}

Algorithm \ref{alg:cloud_verif} details the multi-stage verification process at the Director Repository. Upon receiving the VVM, the repository identifies the vehicle via its VIN and retrieves the corresponding public keys from the inventory database. The process enforces a two-tier cryptographic check: first validating the global VVM signature to ensure the Primary ECU's integrity, then sequentially verifying each secondary ECU's EVR. 

\begin{algorithm}[H]
\caption{Vehicle and Metadata Verification}
\label{alg:cloud_verif}
\begin{algorithmic}[1]
\REQUIRE $VIN$, $VVM$, $DB_{inventory}$, $PR_{dir}$
\ENSURE Vehicle authentication and update target identification.

\STATE \textbf{Receive:} $DR_{vvm} \leftarrow \{VVM(PLOAD_{vvm}, SIG_{attr})\}$
\STATE \textbf{Decode:} $VIN \leftarrow Extract(PLOAD_{vvm})$
\STATE \textbf{Query DB:} $RECORDS_{ecu} \leftarrow DB.Query(VIN)$ \COMMENT{Retrieve keys and image records}
\STATE \textbf{Verify VVM Signature:} 
\STATE $SIGVERI_{vvm} \leftarrow DSVerify(PU_{pecu}, VVM)$
\IF{$SIGVERI_{vvm} == TRUE$}
    \STATE Proceed to ECU validation.
\ELSE
    \STATE \textbf{Abort:} Drop request, log alert.
\ENDIF
\STATE \textbf{Verify individual EVR Signatures:}
\FOR{each $EVR \in VVM$}
    \STATE $SIGVERI_{evr} \leftarrow DSVerify(PU_{ecu}, EVR)$
    \IF{$SIGVERI_{evr} == FALSE$}
        \STATE \textbf{Abort:} Drop request, log alert.
    \ENDIF
\ENDFOR
\STATE \textbf{Return:} \texttt{AUTHENTICATION\_SUCCESS}
\end{algorithmic}
\end{algorithm}

\end{document}